\documentstyle[12pt,epsf]{article}

\begin{document}
 
\newcommand{\be}{\begin{eqnarray}}
\newcommand{\ee}{\end{eqnarray}}

\begin{flushright}
SLAC--PUB--95--7068\\
LBL--37666\\
\end{flushright}
\vspace{1cm}
 
\begin{center}
 
{\Large 
{Charmed Hadron Asymmetries in the Intrinsic Charm Coalescence Model
\footnote{
This work was supported in part by the Director, Office of Energy
Research, Division of Nuclear Physics of the Office of High Energy
and Nuclear Physics of the U. S. Department of Energy under Contract
Numbers DE-AC03-76SF0098 and DE-AC03-76SF00515.} 
}} \\[8ex]
 
R. Vogt\\[2ex]
 Nuclear Science Division\\
 Lawrence Berkeley Laboratory\\
 Berkeley, California\quad 94720 \\
 and\\
 Physics Department\\
 University of California at Davis\\
 Davis, California\quad 95616  \\[2ex]
 
and \\[2ex]
 
S. J. Brodsky
\\[2ex]
 
Stanford Linear Accelerator Center\\
Stanford University\\
Stanford, California\quad 94309  \\[12ex]

(Submitted to Nuclear Physics B.)
\\[3ex]

\end{center}

\newpage
\baselineskip=22pt

\vspace*{0.9cm}
\begin{center}
{\bf ABSTRACT}
\end{center}

Fermilab experiment E791, measuring charmed hadron production in
$\pi^- A$ interactions at 500 GeV with high statistics, has observed
a strong asymmetry  between the hadroproduction cross sections for
leading $D$ mesons which contain projectile valence quarks and  the
nonleading charmed mesons without projectile valence quarks.  Such
correlations of the charge of the $D$ meson with the quantum numbers
of the beam hadron  explicitly contradict the factorization theorem
in perturbative QCD which predicts that heavy quarks hadronize
through  a jet fragmentation function that is independent of the
initial state. The E791 experiment also measures $\Lambda_c/
\overline \Lambda_c$ and $D_s/\overline{D_s}$ production asymmetries
as well as asymmetries in $D \overline D$ pair production. We
examine these asymmetries and the fractional longitudinal momentum,
$x_F$, distributions for single and pairs of charmed hadrons within
a two-component model combining leading-twist $g g$ and $q \overline
q$ fusion subprocesses with charm production from intrinsic heavy
quark Fock states. A key feature of this analysis is intrinsic charm
coalescence, the process by which a charmed quark in the
projectile's Fock state wavefunction forms charmed hadrons by
combining with valence quarks of similar rapidities.
\bigskip
\newpage

\begin{center}
{\bf 1. Introduction}
\end{center}
 
The E791 experiment, studying 500 GeV $\pi^- A$ interactions with
carbon and platinum targets, employs an open geometry spectrometer
with a very open trigger and a fast data acquisition system to
record the world's largest sample of hadroproduced charm
\cite{Carter}.  This large data set allows detailed investigations
of charmed hadron production including $\Lambda_c (udc)$, $\overline
\Lambda_c$, $D_s (c \overline s)$, $\overline D_s$, and $D \overline
D$ pairs.
 
One of the most striking features of charm hadroproduction is the
leading particle effect:  the strong correlation between the quantum
number of the incident hadron and the quantum numbers of the
final-state charmed hadron. For example, more $D^-$  than $D^+$ are
produced at large $x_F$ in $\pi^- A \rightarrow D^\pm X$
\cite{Carter,Agb,Bar,WA82,E7692}. There is also evidence of leading
particle correlations in $\Lambda_c$ \cite{Chauv,Zicc,BIS2} and
$\Lambda_b(udb)$ \cite{Zicc2} production in $pp$ collisions and
$\Xi_c(usc)$ production in hyperon-nucleon interactions
\cite{Biagi,WA89}.  Such correlations are remarkable because they
explicitly contradict the factorization theorem in perturbative QCD
which predicts that heavy quarks hadronize through  a jet
fragmentation function that is independent of the initial state.
 
Leading charm production can be quantified by studies of the
production asymmetries between leading and nonleading charm
production.  In $\pi^- p$ interactions, both  $D^- (\overline c d)$
and $D^0 (c \overline u)$, which share valence quarks with the
$\pi^- (\overline u d)$, are ``leading" while  $D^+ (c \overline d)$
and $\overline{D^0} (\overline c u)$, which do not, are
``nonleading" at $x_F > 0$. The harder leading $D$ distributions
suggest that hadronization at large $x_F$ involves the recombination
of the charmed or anticharmed quarks with the projectile spectator
valence quarks. The $D^-/D^+$ asymmetry is defined as 
\be 
{\cal A}_{D^-/D^+} = \frac{d\sigma(D^-) - d\sigma(D^+)}{d\sigma(D^-)
+ d\sigma(D^+)} \, \, . 
\ee 
The measured asymmetry increases from nearly zero at small $x_F$ to
${\cal A}_{D^-/D^+} \sim 0.5$ around $x_F = 0.65$ \cite{WA82,E7692},
indicating that the leading charm asymmetry is primarily localized
at large $x_F$. Thus the asymmetry ${\cal A}_{D^-/D^+}$ reflects the
physics of only a small fraction of the total $D^\pm$ cross section.
The neutral $D$'s were not used in the analysis since these states
can also be produced indirectly by the decay of nonleading $D^{\star
+}$ mesons.

In a recent paper \cite{VB}, we discussed a QCD mechanism which
produces a strong asymmetry between leading and nonleading charm at
large $x_F.$ A key feature of this analysis is coalescence, the
process by which a produced charmed quark forms charmed hadrons by
combining with quarks of similar rapidities. In leading-twist QCD,
heavy quarks are produced by the fusion subprocesses $g g
\rightarrow Q \overline Q$ and $q \overline q \rightarrow Q
\overline Q$.   The factorization theorem \cite{fact} predicts that
fragmentation is independent of the quantum numbers of both the
projectile and target.  Thus one expects ${\cal A}=0$ to leading
order.  However, it is possible that the forward-moving heavy quarks
will coalescence with the spectator valence quarks of the projectile
to produce leading hadrons in the final state. In a gauge theory
one expects the strongest attraction to occur when the spectator and
produced quarks have equal velocities \cite{BGS}.  Thus the
coalescence probability should be largest at small relative rapidity
and relatively low transverse momentum where the invariant mass  of
the $\overline Q q $ system is minimal, and its amplitude for
binding is maximal.
 
The coalescence of charmed quarks with projectile valence quarks may
also occur in the initial state.  For example, the $\pi^-$ or $p$
wavefunctions can fluctuate into $| \overline u d c \overline c
\rangle$ or $|uu d c \overline c \rangle$ Fock states.  These states
are produced in QCD from amplitudes involving two or more gluons
attached to the charmed quarks.  The most important fluctuations
occur at minimum invariant mass $M$ where all the partons have
approximately the same velocity. These fluctuations can have very
long lifetimes in the target rest frame, ${\cal O}( 2 P_{\rm
lab}/M^2)$, where $P_{\rm lab}$ is the projectile momentum. Since
the charm and valence quarks have the same rapidity in these states,
the heavy quarks carry a large fraction of the projectile momentum.
Furthermore the comoving heavy and light quarks can readily coalesce
to produce leading charm correlations at a large combined
longitudinal momentum. Such a mechanism can dominate the
hadroproduction rate at large $x_F$. This is the underlying
assumption of the intrinsic charm model \cite{intc}.

The intrinsic charm fluctuations in the wavefunction are initially
far off the light-cone energy shell shell. However, they become on
shell and materialize into charmed hadron when a light spectator
quark in the projectile  Fock state interacts  in the target
\cite{BHMT}. Since such interactions are strong, the charm
production will occur primarily on the front face of the nucleus in
the case of a nuclear target.  Thus an important characteristic of
the intrinsic charm model is its strong nuclear dependence; the
cross section for charm production via the materialization of heavy
Fock states should have a nuclear dependence at high energies
similar to that of inelastic hadron-nucleus cross sections.
 
In this work,  we concentrate on the charmed hadrons and meson pairs
channels studied by E791 in order to further examine the
relationship between fragmentation and coalescence mechanisms. The
calculations are made within a two-component model: leading-twist
fusion and intrinsic charm \cite{VB,VBH1,VBH2}. We find that the
coalescence of the intrinsic charmed quarks with the valence quarks
of the projectile is the dominant mechanism for producing fast $D$
and $D^\star$ mesons. On the other hand, when the charmed quarks
coalesce with sea quarks, there is no leading charmed hadron. We
discuss the longitudinal momentum distributions and the related
asymmetries for  $\Lambda_c/\overline \Lambda_c$ and $D_s/\overline
D_s$ production,  as well as $D \overline D$ pairs. (We have only
applied our model to $D \overline D$ pairs in the forward hemisphere
in order to provide a clear definition of the asymmetry.)
 
As expected, the asymmetries predicted by the intrinsic charm
coalescence model are a strong function of $x_F$. We find that
$\Lambda_c$ production in the proton fragmentation region ($x_F <0$
in $\pi^- p$ collisions) is dominated by the coalescence of the
intrinsic charm quark with the $u d$ valence quarks of the proton.
Coalescence is particularly important in $D \overline D$ pair
production.  The production of $D_s/\overline D_s$ and, at $x_F>0$,
$\Lambda_c/\overline \Lambda_c$ by coalescence must occur within
still higher particle number Fock states.
   
Leading particle correlations are also an integral part  of the
Monte Carlo program PYTHIA \cite{PYT} based on the Lund string
fragmentation model. In this model it is assumed that the heavy
quarks are produced in the initial state with relatively small
longitudinal momentum fractions by the leading twist fusion
processes. In order to produce a strong leading particle effect at
large $x_F$, the string has to accelerate the heavy quark as it
fragments and forms the final-state heavy hadron.  Such a mechanism
goes well beyond the usual assumptions made in hadronization models
and arguments based on heavy quark symmetry, since it demands that 
large changes of the heavy quark momentum take place in the final
state.

In this paper we shall compare the  predictions of the intrinsic
charm coalscence model with those of PYTHIA \cite{PYT}. The
comparison of the data with these models  should distinguish the
importance of higher heavy quark Fock state fluctuations in the
initial state and the coalescence process from the strong string
hadronization effects postulated in the PYTHIA model.\\[3ex]
 
\begin{center}
{\bf 2. Leading-Twist Production}
\end{center}
 
In this section we briefly review the conventional leading-twist
model for the production of single charmed hadrons and $D \overline
D$ pairs in $\pi^- p$ interactions. We will also show the
corresponding distributions of charmed hadrons predicted by the 
PYTHIA model \cite{PYT}.
 
Our calculations are at lowest order in $\alpha_s$.  A constant
factor $K \sim 2-3$ is included in the fusion cross section since
the next-to-leading order $x_F$ distribution is larger than the
leading order distribution by an approximately constant factor
\cite{RV2}. Neither leading order production nor the next-to-leading
order corrections can produce flavor correlations \cite{Frix}.

The single charmed hadron $x_F$ distribution, $x_F = (2m_T/\sqrt{s})
\sinh y$, has the factorized form \cite{VBH2}
\be
\frac{d\sigma}{dx_F} = \frac{\sqrt{s}}{2} \int H_{ab}(x_a,x_b)
\frac{1}{E_1}\ \frac{D_{H/c}(z_3)}{z_3}\ dz_3\, dy_2\, dp_T^2 \, \, ,
\ee
where $a$ and $b$ are the initial partons, 1 and 2 are the charmed
quarks with $m_c = 1.5$ GeV, and 3 and 4 are the charmed hadrons.
The convolution of the subprocess cross sections for $q \overline q$
annihilation and gluon fusion with the parton densities is included
in $H_{ab} (x_a, x_b)$,
\be 
H_{ab}(x_a,x_b) =  \sum_q [f_q^A(x_a)  f_{\overline q}^B(x_b) +
f_{\overline q}^A(x_a) f_q^B(x_b)] \frac{d \widehat{\sigma}_{q
\overline q}}{d \hat{t}} + f_g^A(x_a) f_g^B(x_b) \frac{d
\widehat{\sigma}_{gg}}{d \hat{t}} \, \, ,
\ee
where $A$ and $B$ are the interacting hadrons. For consistency with
the leading-order calculation, we use current leading order parton
distribution functions, GRV LO, for both the nucleon \cite{GRV} and
the pion \cite{GRVpi}.
 
The fragmentation functions, $D_{H/c}(z)$, describe the
hadronization of the charmed quark where $z = x_H/x_c$ is the
fraction of the charmed quark momentum carried by the charmed
hadron, produced roughly collinear to the charmed quark. Assuming
factorization, the fragmentation is independent of the initial state
(leptons or hadrons) and thus cannot produce flavor correlations
between the projectile valence quarks and the charmed hadrons. This
uncorrelated fragmentation will be modeled by two extremes: a delta
function, $\delta(z -1)$, and the Peterson function \cite{Pete}, as
extracted from $e^+e^-$ data. The Peterson function predicts a
softer $x_F$ distribution than observed in hadroproduction, even at
moderate $x_F$ \cite{VBH2}, since the fragmentation decelerates the
charmed quark, decreasing its average momentum fraction, $\langle
x_F \rangle$, approximately 30\%\ by the production of $D$ mesons. 
The delta-function model assumes that the  charmed quark coalesces
with a low-$x$ spectator sea quark or a low momentum secondary quark
such that the charmed quark retains its momentum \cite{VBH2}. This
model is more consistent with low $p_T$ charmed hadroproduction data
\cite{AgB1,AgB2,Bar2} than Peterson fragmentation.
 
The parameters of the Peterson function we use here are taken from
$e^+ e^-$ studies of $D$ production \cite{Chirn}. The $D^\star$
distributions are very similar to the $D$ distributions but the
$\Lambda_c$ distribution appears to be somewhat softer
\cite{Cashmore}. Although there is some uncertainty in the exact
form of the Peterson function for charmed baryons and $D_s$ mesons,
it always produces deceleration.

\vspace{.5cm}
\begin{figure}[htbp]
\begin{center}
\leavevmode
\epsfbox{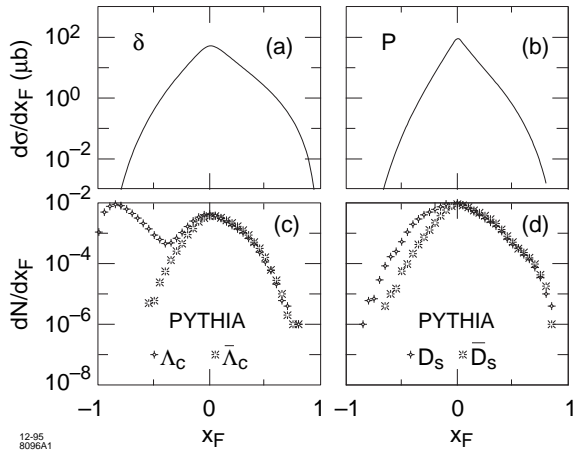}
\end{center}
\caption[*]{Leading-twist fusion calculations of single charm
production from $\pi^- p$ interactions at 500 GeV with delta
function (a) and Peterson function (b) fragmentation.  The results
from the PYTHIA Monte Carlo \cite{PYT} for $\Lambda_c/\overline
\Lambda_c$ (c) and $D_s/\overline D_s$ (d) production are also
shown.  The fusion calculations are normalized to the charm cross
section while the PYTHIA distributions are normalized to the number
per event. } 
\label{fig1} 
\end{figure}

In Fig.\ 1 we show the single inclusive $x_F$ distributions
calculated for (a) delta function and (b) Peterson function
fragmentation in $\pi^- p$ interactions at 500 GeV.  The results are
normalized to the total single charmed quark cross section.  The
parton distributions of the pion are harder than those of the proton
at large $x_{\rm Bj}$, producing broader forward distributions.  As
expected, the delta-function fragmentation results in harder
distributions than those predicted by Peterson fragmentation for
$x_F > 0.2$. However, as shown in \cite{WA82}, the conventional
fusion model, even with delta-function fragmentation, cannot account
for the shape of leading $D$ distributions.
 
The charmed hadron distributions from PYTHIA, obtained from a $\pi^-
p$ run with $10^6$ events at 500 GeV using all default settings and
the GRV LO parton distributions, are shown in Fig.\ 1(c) for
$\Lambda_c$ and $\overline \Lambda_c$ and 1(d) for $D_s$ and
$\overline D_s$ hadrons. The distributions are normalized to the
number of charmed hadrons per event.  PYTHIA is based on the Lund
string fragmentation model \cite{PYT} in which charmed quarks are at
string endpoints.  The strings pull the charmed quarks toward the
opposite string endpoints, usually beam remnants.  When the two
string endpoints are moving in the same general direction, the
charmed hadron can be produced with larger longitudinal momentum
than the charmed quark, accelerating it.  In the extreme case where
the string invariant mass is too small to allow multiple particle
production, this picture reduces to final-state coalescence and the
string endpoints determine the energy, mass, and flavor content of
the produced hadron \cite{torb}. Thus a $D^-$ or $D^0$ can inherit
all of the remaining projectile momentum while $D^+$ and
$\overline{D^0}$ production is forbidden. The coalescence of a
charmed quark with a valence diquark results in the secondary peak
at $x_F \approx -0.85$ in the $\Lambda_c$ distribution shown in
Fig.\ 1(c).  The $\Lambda_c$ baryon is a leading charmed hadron in
the proton fragmentation region since it can have two valence quarks
in common with the proton.  Also in the proton fragmentation region,
the $D_s$ is somewhat harder than the $\overline D_s$.  This is
evidently a secondary effect of $\Lambda (uds)$ coalescence with a
$ud$ diquark: the $\overline s$ can pull the charmed quark to larger
$x_F$ in the wake of the $\Lambda$, producing more $D_s$ at negative
$x_F$ than $\overline D_s$. Such coalescence correlations are not
predicted for the $\overline \Lambda_c$ or the $\overline D_s$ in
the proton fragmentation region.  In the pion fragmentation region,
the predicted $D_s/\overline D_s$ distributions are harder than the
$\Lambda_c/\overline \Lambda_c$ distributions.
 
The charmed pair $x_F$ distribution is
\be 
\frac{d \sigma}{dx_{F}} & = & \int H_{ab}(x_a,x_b)
\frac{E_3E_4}{E_1E_2}\ \frac{D_{H/c}(z_3)D_{\overline H/\overline
c}(z_4)}{z_3z_4} \ \delta(M_{D \overline D}^2 - 2m_T^2(1+\cosh
(y_3-y_4))) \nonumber \\  &   & \mbox{} \times \delta(x_{F} - x_3 -
x_4) \, dz_3 dz_4  dy_3 dy_4 dp_T^2 dM_{D \overline D}^2 \, \, , 
\ee
where $x_{F} = x_3 + x_4$ and $m_T^2 = p_T^2 + m_D^2$. Figure 2
shows the forward $x_F$ distribution of $D \overline D$ pairs from
$\pi^- p$ interactions at 500 GeV with (a) delta function and (b)
Peterson function fragmentation for both charmed quarks.  The
distributions are normalized to the total $c \overline c$ pair
production cross section. Note that the pair distributions are
harder than the single distributions in Fig.\ 1.  The perturbative
QCD calculation cannot distinguish between leading and nonleading
hadrons in the pair distributions.

\vspace{.5cm}
\begin{figure}[htbp]
\begin{center}
\leavevmode
\epsfbox{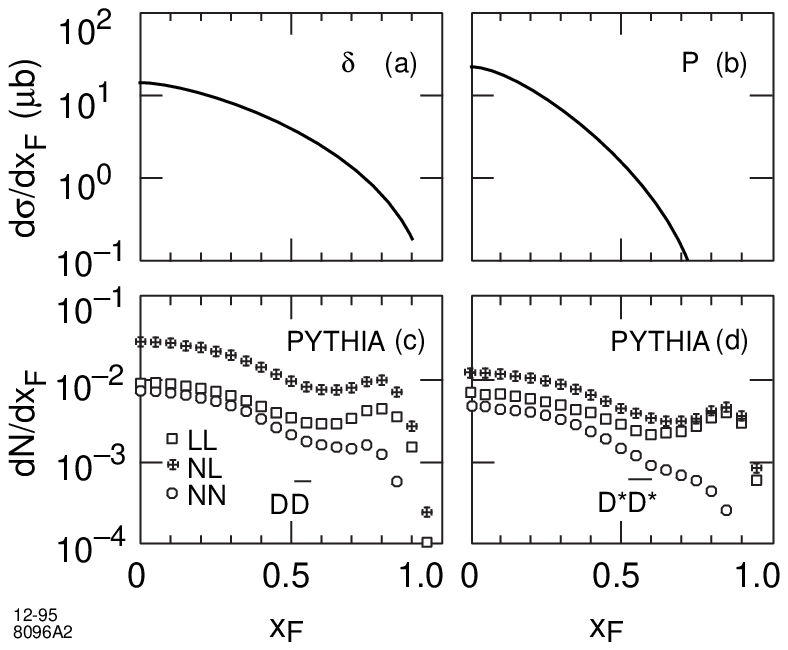}
\end{center}
\caption[*]{Leading-twist fusion calculations of $c \overline c$
pair production from $\pi^- p$ interactions at 500 GeV with delta
function (a) and Peterson function (b) fragmentation.  The results
from the PYTHIA Monte Carlo \cite{PYT} for $LL$, $NL$, and $NN$ $D
\overline D$ (c) and $D^\star \overline D^\star$ (d) pairs are also
shown.  The fusion calculations are normalized to the $c \overline
c$ cross section while the PYTHIA distributions are normalized to
the number per event.
}
\label{fig2}
\end{figure}
 
In Fig.\ 2(c), the $D \overline D$ pair distributions from PYTHIA
have been classified as doubly leading $(LL)$,  $D^-D^0$,
nonleading-leading $(NL)$,  $D^-D^+ + D^0\overline{D^0}$, and doubly
nonleading $(NN)$, $D^+ \overline{D^0}$.  The same classification
for $D^\star$ pairs is shown in Fig.\ 2(d).  The distributions are
normalized to the number of charmed pairs per event. We have not
considered $D \overline{D^\star}$ or $D^\star \overline D$ pairs. 
Note that the leading particle assignments are only valid for $x_F >
0$.  The assignments are more meaningful for the $D^\star$'s since
they may be assumed to be directly produced.  The neutral $D$ mesons
in Fig.\ 2(c) arise in part from charged $D^\star$ decays. Thus the
$NN$ $D$ pair distributions have a shoulder at $x_F \approx 0.8$
from $D^{\star -}$ decays to $\overline{D^0}$ which is absent in the
$NN$ $D^\star \overline{D^\star}$ distributions.  Note also that the
$NL$ pairs are most numerous since both charged and neutral $D$
pairs contribute to the distribution.  Almost three times as many
neutral $D$'s are produced than charged $D$'s.  Only some of this
difference can arise from $D^\star$ decays since charged and neutral
$D^\star$'s are produced in nearly equal abundance.
 
Other final-state coalescence models have been proposed, including a
valence spectator recombination model \cite{bedR} and the valon
model \cite{hwa}. Two important unknowns in these models are the
correlation between the charmed quark and the valence spectator in
the recombination function and the $n$-particle parton distributions
of the spectator and participant valence quarks. In this work we
will not compare to either of these models but simply note that they
can also produce charmed hadrons and hadron pairs by final-state
coalescence.\\[3ex]
 
\begin{center}
{\bf 3.  Intrinsic Heavy Quark Production}
\end{center}
 
The wavefunction of a hadron in QCD can be represented as a
superposition of Fock state fluctuations, {\it e.g.}\ $\vert n_V
\rangle$, $\vert n_V g \rangle$, $\vert n_V Q \overline Q \rangle$,
\ldots components where $n_V \equiv \overline u d$ for a $\pi^-$ and
$uud$ for a proton. When the projectile scatters in the target, the
coherence of the Fock components is broken and the fluctuations can
hadronize either by uncorrelated fragmentation or coalescence with
spectator quarks in the wavefunction \cite{intc,BHMT}.  The
intrinsic heavy quark Fock components are generated by virtual
interactions such as $g g \rightarrow Q \overline Q$ where the
gluons couple to two or more projectile valence quarks. The
probability to produce $Q \overline Q$ fluctuations scales as
$\alpha_s^2(m_{Q \overline Q})/m_Q^2$ relative to leading-twist
production \cite{BH} and is thus higher twist. Intrinsic $Q
\overline Q$ Fock components are dominated by configurations with
equal rapidity constituents so that, unlike sea quarks generated
from a single parton, the intrinsic heavy quarks carry a large
fraction of the parent momentum \cite{intc}.
 
The frame-independent probability distribution of an $n$--particle
$c \overline c$ Fock state is 
\be 
\frac{dP_{\rm ic}}{dx_i \cdots dx_n} = N_n \alpha_s^4(M_{c \overline
c}) \frac{\delta(1-\sum_{i=1}^n x_i)}{(m_h^2 - \sum_{i=1}^n
(\widehat{m}_i^2/x_i) )^2} \, \, , 
\ee 
where $N_n$, assumed to be slowly varying, normalizes the $|n_V c
\overline c \rangle$ probability, $P_{\rm ic}$.  The delta function
conserves longitudinal momentum.  The dominant Fock configurations
are closest to the light-cone energy shell shell and therefore have
minimal invariant mass, $M^2 = \sum_i \widehat{m}_i^2/ x_i$, where
$\widehat{m}_i^2 =k^2_{T,i}+m^2_i$ is the effective transverse mass
of the $i^{\rm th}$ particle and $x_i$ is the light-cone momentum
fraction.  Assuming $\langle \vec k_{T, i}^2 \rangle$ is
proportional to the square of the constituent quark mass, we adopt
the effective values $\widehat{m}_q = 0.45$ GeV, $\widehat{m}_s =
0.71$ GeV, and $\widehat{m}_c = 1.8$ GeV \cite{VBH1,VBH2}.
 
The intrinsic charm production cross section can be related to
$P_{\rm ic}$ and the inelastic $hN$ cross section by 
\be 
\sigma_{\rm ic}(hN) = P_{\rm ic} \sigma_{h N}^{\rm in}
\frac{\mu^2}{4 \widehat{m}_c^2} \, \, . 
\ee 
The factor of $\mu^2/4 \widehat{m}_c^2$ arises because a soft
interaction is needed to break the coherence of the Fock state. The
NA3 collaboration \cite{Badier} separated the nuclear dependence of
$J/\psi$ production in $\pi^- A$ interactions into a ``hard"
contribution with a nearly linear $A$ dependence at low $x_F$ and a
high $x_F$ ``diffractive" contribution scaling as $A^{0.77}$,
characteristic of soft interactions. One can fix the soft
interaction scale parameter, $\mu^2 \sim 0.2$ GeV$^2$, by the
assumption that the diffractive fraction of the total production
cross section \cite{Badier} is the same for charmonium and charmed
hadrons.  Therefore, we obtain $\sigma_{\rm ic}(\pi N) \approx 0.5$
$\mu$b at 200 GeV and $\sigma_{\rm ic}(p N) \approx 0.7$ $\mu$b
\cite{VB} with $P_{\rm ic} = 0.3$\%\ from an analysis of the EMC
charm structure function data \cite{EMCic}. A recent reanalysis of
this data with next-to-leading order calculations of leading twist
and intrinsic charm electroproduction confirms the presence of an
$\approx 1$\%\ intrinsic charm component in the proton for large
$x_{\rm Bj}$ \cite{hsv}.  Note that a larger $P_{\rm ic}$ would not
necessarily lead to a larger $\sigma_{\rm ic}$. Since we have fixed
$\mu^2$ from the NA3 data, increasing $P_{\rm ic}$ would decrease
$\mu^2$ accordingly.
 
We now calculate the probability distributions, $dP_{\rm ic}/dx_F$
for charmed hadrons and $D \overline D$ pairs resulting from both
uncorrelated fragmentation and coalescence of the quarks in the
intrinsic charmed Fock states.  These light-cone distributions are
frame independent.  In a hadronic interaction, these states are
dissociated and materialize with the corresponding differential
cross section  
\be 
\frac{d\sigma_{\rm ic}(hN)}{dx_F} = \sigma_{h N}^{\rm in}
\frac{\mu^2}{4 \widehat{m}_c^2} \frac{dP_{\rm ic}}{dx_F}\, \, . 
\ee 
In the case of $\pi^- p$ collisions, the fluctuations of the
$|\overline u d c \overline c \rangle$ state produces charmed
hadrons at $x_F > 0$ in the center of mass while the fluctuations of
the $|uud c \overline c \rangle$ state produces charmed hadrons at
$x_F < 0$.\\[1ex]
 
\noindent
{\bf 3.1  Single charmed hadrons}
\medskip

There are two ways of producing charmed hadrons from intrinsic $c
\overline c$ states.  The first is by uncorrelated fragmentation,
discussed in Section 2.  Additionally, if the projectile has the
corresponding valence quarks, the charmed quark can also hadronize
by coalescence with the valence spectators. The coalescence
mechanism thus introduces flavor correlations between the projectile
and the final-state hadrons, producing {\it e.g.}\ $D^-$'s with a
large fraction of the $\pi^-$ momentum.  In the pion fragmentation
region, $x_F >0$, $D^-$ and $D^0$ have contributions from both
coalescence and fragmentation while $D^+$ and $\overline{D^0}$ can
only be produced from the minimal $c \overline c$ Fock state by
fragmentation. In the proton fragmentation region, $x_F <0$, the
$D^-$ and $\overline{D^0}$ may be produced by both coalescence and
fragmentation.
 
If we assume that the $c$ quark fragments into a $D$ meson, the $D$
distribution is 
\be 
\frac{d P^F_{\rm ic}}{dx_{D}} = \int dz \prod_{i=1}^n dx_i
\frac{dP_{\rm ic}}{dx_1 \ldots dx_n} D_{D/c}(z) \delta(x_{D} - z
x_c) \, \, , 
\ee 
where  $n=4$, 5 for pion and proton projectiles in the $|n_V c
\overline c \rangle$ configuration. This mechanism produces $D$
mesons carrying 25-30\%\ of the projectile momentum with the delta
function and 17-20\%\ with the Peterson function. The $D$
distributions are shown in Fig.\ 3(a) and 3(b) for proton and pion
projectiles normalized to the total probability of the $|n_V c
\overline c \rangle$ Fock state configuration assuming that $P_{\rm
ic}$ is the same for protons and pions.  Less momentum is given to
the charmed quarks in the proton than in the pion because the total
momentum is distributed among more partons.   These distributions
are assumed for all intrinsic charm production by uncorrelated
fragmentation.
 
\vspace{.5cm}
\begin{figure}[htbp]
\begin{center}
\leavevmode
\epsfbox{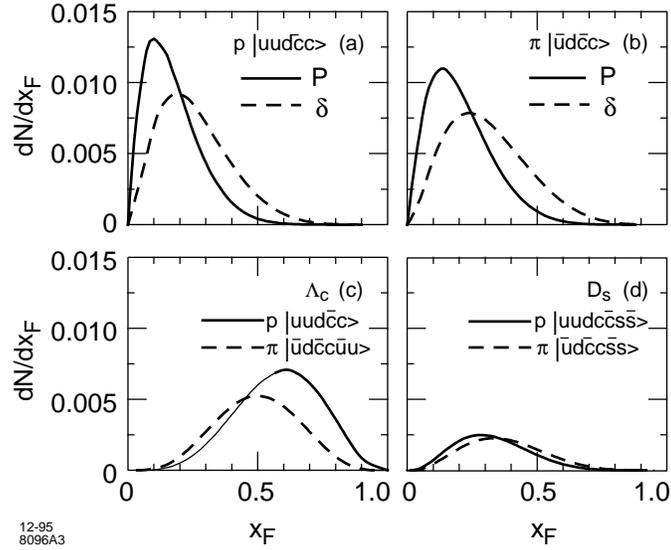}
\end{center}
\caption[*]{Charmed particle distributions from the intrinsic charm
model.  The charmed quark fragmentation distributions are shown for
Peterson function (solid) and delta function (dashed) fragmentation
for proton (a) and pion (b) projectiles.  Charmed hadrons produced
by coalescence from protons (solid) and pions (dashed) are also
shown for (c) $\Lambda_c$ baryons and (d) $D_s$ mesons.  The charmed
quark distributions in (a) and (b) and the $\Lambda_c$ distribution
from a proton (c) are normalized to $P_{\rm ic}$.  The $\Lambda_c$
distribution from a pion (c) and the $D_s$ distributions (d) are
normalized to $P_{\rm icu}$ and $P_{\rm ics}$ respectively.
}
\label{fig3}
\end{figure}

The coalescence distributions, on the other hand, are specific for
the individual charmed hadrons.  The coalescence contribution to
leading $D$ production is 
\be 
\frac{d P^C_{\rm ic}}{dx_{D}} = \int \prod_{i=1}^n dx_i
\frac{dP_{\rm ic}}{dx_1 \ldots dx_n} \delta(x_{D} - x_c - x_1) \, \,
. 
\ee  
With the additional momentum of the light valence quark, the $D$
takes 40-50\%\ of the momentum. In the proton fragmentation region,
the $c$ quark can coalesce with valence $u$ and $d$ quarks to
produce leading $\Lambda_c$'s, 
\be 
\frac{d P^C_{\rm ic}}{dx_{\Lambda_c}} = \int \prod_{i=1}^n dx_i
\frac{dP_{\rm ic}}{dx_1 \ldots dx_n} \delta(x_{\Lambda_c} - x_c -
x_1 - x_2) \, \, , 
\ee 
carrying 60\%\ of the proton momentum.  The distribution, shown in
Fig.\ 3(c), is also normalized to $P_{\rm ic}$.
 
Coalescence may also occur within higher fluctuations of the
intrinsic charm Fock state.  For example, at $x_F >0$, $\overline
\Lambda_c$ and $D_s$ can be produced by coalescence from $|n_V c
\overline c d \overline d \rangle$ and $|n_V c \overline c s
\overline s \rangle$ configurations.  We previously studied $\psi
\psi$ production from  $|n_V c \overline c c \overline c \rangle$
states \cite{VB2}. Assuming that all the measured $\psi \psi$ pairs
\cite{Badpi,Badp} arise from these configurations, we can relate the
$\psi \psi$ cross section, 
\be 
\sigma_{\rm ic}^{\psi \psi} (hN) = f_{\psi/h}^2 \frac{P_{\rm
icc}}{P_{\rm ic}} \sigma_{\rm ic} (hN)  \, \, , 
\ee 
to the double intrinsic charm production probability, $P_{\rm icc}$,
where $f_{\psi/h}$ is the fraction of intrinsic $c \overline c$
pairs that become $J/\psi$'s.  The upper bound on the model,
$\sigma_{\psi \psi} = \sigma_{\rm ic}^{\psi \psi} (\pi^- N) \approx
20$ pb \cite{Badpi}, requires $P_{\rm icc} \approx 4.4\%\ P_{\rm
ic}$ \cite{VB2,RV}. This value of $P_{\rm icc}$ can be used to
estimate the probability of light quark pairs in an intrinsic charm
state. We expect that the probability of additional light quark
pairs in the Fock states to be larger than $P_{\rm icc}$, 
\be 
P_{\rm icq} \approx \left( \frac{\widehat{m}_c}{\widehat{m}_q}
\right)^2 P_{\rm icc} \, \, , 
\ee 
leading to $P_{\rm icu} = P_{\rm icd} \approx 70.4\%\ P_{\rm ic}$
and $P_{\rm ics} \approx 28.5\%\ P_{\rm ic}$.
 
Then with a pion projectile at $x_F >0$, the $\overline \Lambda_c$
coalescence distribution from a six-particle Fock state is 
\be
\frac{d P^C_{\rm ic}}{dx_{\overline \Lambda_c}} = \int \prod_{i=1}^n
dx_i \frac{dP_{\rm ic}}{dx_1 \ldots dx_n} \delta(x_{\overline
\Lambda_c} - x_{\overline c} - x_1 - x_{\overline d}) \, \, , 
\ee 
also shown in Fig.\ 3(c) and normalized to $P_{\rm icu}$.  Since
half of the quarks are needed to produce the $\overline \Lambda_c$,
it carries 50\%\ of the pion momentum.  The $D_s$ mesons arising
from coalescence in the $|n_V c \overline c s \overline s \rangle$
state, 
\be 
\frac{d P^C_{\rm ic}}{dx_{D_s}} = \int \prod_{i=1}^n dx_i
\frac{dP_{\rm ic}}{dx_1 \ldots dx_n} \delta(x_{D_s} - x_c -
x_{\overline s}) \, \, , 
\ee 
carry $\approx 30$-40\%\ of the hadron momentum, as shown in Fig.\
3(d) and normalized to $P_{\rm ics}$. The $D_s$ and $\overline D_s$
inherit less total momentum than the leading $D$ since the Fock
state momentum is distributed over more partons.  Thus as more
partons are included in the Fock state, the coalescence
distributions soften and approach the fragmentation distributions,
Eq.\ (8), eventually producing charmed hadrons with less momentum
than uncorrelated fragmentation from the minimal $c \overline c$
state if a sufficient number of $q \overline q$ pairs are included. 
There is then no longer any advantage to introducing more light
quark pairs into the configuration--the relative probability will
decrease while the potential gain in momentum is not significant. 
We thus do not consider $\overline \Lambda_c$ production by
coalescence at $x_F <0$ since a minimal nine-parton Fock state is
required.\\[1ex]
 
\noindent
{\bf 3.2  $D \overline D$ pair production}
\medskip
 
In any $|n_V c \overline c \rangle$ state, $D \overline D$ pairs may
be produced by double fragmentation, a combination of fragmentation
and coalescence, or, from a pion projectile only, double
coalescence.  We discuss only pair production at $x_F >0$ so that
$|\overline u d c \overline c \rangle$ is the minimal Fock state. 
In the proton fragmentation region, with no valence antiquark, the
leading mesons would be $D^- (\overline c d)$ and $\overline{D^0}
(\overline c u)$.  Therefore no doubly leading $D^-
\overline{D^0}$ pairs can be produced from the five parton state: two
intrinsic $c \overline c$ pairs are needed, {\it i.e.}\ $|uud c
\overline c c \overline c \rangle$ states, automatically softening
the effect. However, doubly leading meson-baryon pairs such as
$\Lambda_c \overline{D^0}$ and $\Sigma_c^{++} D^-$ may be produced
by coalescence in the $|uud c \overline c \rangle$ state.  These
combinations might be interesting to study in $pp$ interactions.
 
\vspace{.5cm}
\begin{figure}[htbp]
\begin{center}
\leavevmode
\epsfbox{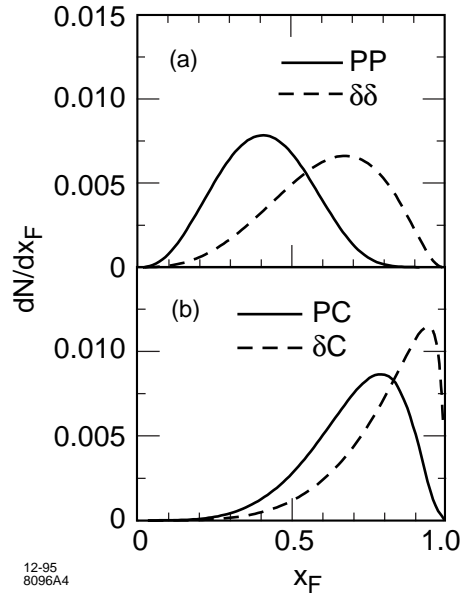}
\end{center}
\caption[*]{Intrinsic $D \overline D$ distributions from a $\pi^-$
projectile. Double fragmentation distributions are shown for
Peterson function (solid) and delta function (dashed) fragmentation
in (a). Charmed pairs produced by coalescence of one quark and
fragmentation of the other are shown in (b) for Peterson (solid) and
delta (dashed) function fragmentation.  The double coalescence
production from a pion is a delta function at $x_F = 1$.  The
distributions are normalized to $P_{\rm ic}$.
}
\label{fig4}
\end{figure}

The $D \overline D$ pairs resulting from double fragmentation, 
\be
\frac{d P^{FF}_{\rm ic}}{dx_{D \overline D}} 
& = & \int dz_c dz_{\overline c} \prod_{i=1}^n dx_i \frac{dP_{\rm
ic}}{dx_1 \ldots dx_n} D_{D/c}(z_c) D_{\overline D/\overline
c}(z_{\overline c}) \\ \nonumber 
&   &  \mbox{} \times \delta(x_{D} - z_c x_c) \delta(x_{\overline D}
- z_{\overline c} x_{\overline c}) \delta(x_{D \overline D} - x_D -
x_{\overline D}) \, \, , 
\ee  
carry the lowest fraction of the projectile momentum.  The
distributions are shown in Fig.\ 4(a). When both charmed quarks
fragment by the Peterson function, $\sim 40$\%\ of the $\pi^-$
momentum is given to the pair while the average is 62\%\ with delta
function fragmentation. If, {\it e.g.}\ a $D$ is produced by
coalescence while the $\overline D$ is produced by fragmentation,
\be 
\frac{d P^{FC}_{\rm ic}}{dx_{D \overline D}} 
& = & \int dz \prod_{i=1}^n dx_i \frac{dP_{\rm ic}}{dx_1 \ldots
dx_n} D_{\overline D/\overline c}(z) \\ \nonumber 
&   & \mbox{}
\times \delta(x_{D} - x_c - x_1) \delta(x_{\overline D} - z
x_{\overline c}) \delta(x_{D \overline D} - x_D - x_{\overline D})
\, \, . 
\ee 
These distributions, with rather large momentum fractions, 71\%\ for
Peterson fragmentation and 87\%\ for the delta function, are shown
in Fig.\ 4(b).  When the projectile has a valence antiquark, as in
the pion, $D \overline D$ pair production by double coalescence is
possible, 
\be 
\frac{d P^{CC}_{\rm ic}}{dx_{D \overline D}} = \int
\prod_{i=1}^n dx_i \frac{dP_{\rm ic}}{dx_1 \ldots dx_n} \delta(x_{D}
- x_c - x_1) \delta(x_{\overline D} - x_{\overline c} - x_2)
\delta(x_{D \overline D} - x_D - x_{\overline D}) \, \, . 
\ee 
All of the momentum of the four-particle Fock state is transferred
to the $D \overline D$ pair, {\it i.e.}\ $x_{D \overline D} \equiv
1$. We also consider double coalescence from a pion in a six
particle Fock state.  In this case, 74\%\ of the pion momentum is
given to the pair, similar to the result for Peterson fragmentation
with coalescence, Eq.\ (16), as could be expected from our
discussion of $D_s$ production in this model.\\[3ex]
 
\begin{center}
{\bf 4.  Predictions of the Two-Component Model}
\end{center}
 
We now turn to specific predictions of the $x_F$ distributions and
asymmetries in our model. The $x_F$ distribution is the sum
of the leading-twist fusion and intrinsic charm components, 
\be
\frac{d\sigma}{dx_F} = \frac{d\sigma_{\rm lt}}{dx_F} +
\frac{d\sigma_{\rm ic}}{dx_F} \, \, , 
\ee 
where $d\sigma_{\rm ic}/dx_F$ is related to $dP_{\rm ic}/dx_F$ in
Eq.\ (7). Note that when we discuss uncorrelated fragmentation, the
same function, either the delta or Peterson function, is used for
both leading twist fusion and intrinsic charm.  The intrinsic charm
model produces charmed hadrons by a mixture of uncorrelated
fragmentation and coalescence \cite{VB,VBH2}. Coalescence in the
intrinsic charm model is taken to enhance the leading charm
probability over nonleading charm.  Since we have not made any
assumptions about how the charmed quarks are distributed into the
final-state charmed hadron channels, an enhancement by coalescence
is not excluded as long as the total probability of all charmed
hadron production by intrinsic charm does not exceed $P_{\rm ic}$. 
Because little experimental guidance is available to help us
separate the charm production channels, we have not directly
addressed the issue here. Thus the above distributions are
normalized to the total $c \overline c$ cross section for the pair
distributions and to the single charm cross section for the single
charmed hadrons.  This is naturally an overestimate of the cross
sections in the charm channels, but more complete measurements are
needed before the relative strengths of the $D$, $\Lambda_c$, $D_s$,
$D^\star$, {\it etc.}\ contributions to the charm cross section can
be understood.
 
In the case of nuclear targets, the model assumes a linear $A$
dependence for leading-twist fusion and an $A^{0.77}$ dependence for
the intrinsic charm component \cite{Badier}.  This $A$ dependence is
included in the calculations of the production asymmetries while the
$x_F$ distributions are given for $\pi^- p$ interactions.  The
intrinsic charm contribution to the longitudinal momentum
distributions is softened if the $A$ dependence is included.\\[1ex]
 
\noindent
{\bf 4.1 Single charmed hadrons}
\medskip
 
We now consider the single charmed hadron distributions produced in
$\pi^- p$ interactions at 500 GeV over the entire $x_F$ range. 
Since the production mechanisms are somewhat different for positive
and negative $x_F$, particularly for the $\Lambda_c$, we will
discuss the pion and proton fragmentation regions separately.
 
We begin with $\Lambda_c$ production in the proton fragmentation
region, negative $x_F$.  As we have already indicated, the $\Lambda_c$
can be produced by coalescence from the $|uudc \overline c
\rangle$ configuration with an average of 60\%\ of the
center-of-mass proton momentum.  The $\overline \Lambda_c$ can only
be produced by fragmentation from a five-particle Fock state and, if
a nine-particle Fock state is considered, the coalescence
distribution will not be significantly harder than the fragmentation
distribution shown in Fig.\ 3(a) since four additional light quarks
are included in the minimal proton Fock state.
 
Therefore coalescence is only important for the $\Lambda_c$, leading
naturally to an asymmetry between $\Lambda_c$ and $\overline
\Lambda_c$. We will assume that the same number of $\Lambda_c$ and
$\overline \Lambda_c$ are produced by fragmentation and that any
excess of $\Lambda_c$ production is solely due to coalescence. 
Then, at $x_F < 0$,
\be 
\frac{dP^{\overline \Lambda_c}_{\rm ic}}
     {dx_{\overline \Lambda_c}} & = &
\frac{dP^F_{\rm ic}}{dx_{\overline \Lambda_c}} \\
\frac{dP^{\Lambda_c}_{\rm ic}}{dx_{\Lambda_c}} & = &
\frac{dP^F_{\rm ic}}{dx_{\Lambda_c}} \, + \, r \,
\frac{dP^C_{\rm ic}}{dx_{\Lambda_c}}
\, \, . 
\ee  
The fragmentation distribution is taken from Eq.\ (8), the
coalescence distribution from Eq.\ (10). The parameter $r$ is
related to the integrated ratio of $\Lambda_c$ to $\overline
\Lambda_c$ production. We have assumed three values of $r$: 1, 10,
and, as an extreme case, 100.  The results for the two uncorrelated
fragmentation functions are shown in Fig.\ 5(a) and 5(b).  Intrinsic
charm fragmentation produces a slight broadening of the $\overline
\Lambda_c$ distribution for delta function fragmentation over
leading-twist fusion (increasing to a shoulder for the Peterson
function).  The $\Lambda_c$ distribution, strongly dependent on $r$,
is considerably broadened.

\vspace{.5cm}
\begin{figure}[htbp]
\begin{center}
\leavevmode
\epsfbox{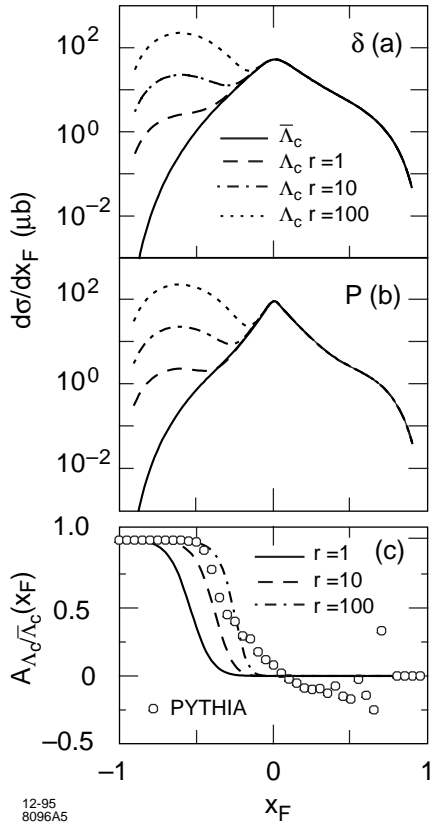}
\end{center}
\caption[*]{The $\Lambda_c/\overline \Lambda_c$ $x_F$ distributions
predicted by the two-component model in (a) and (b) for delta and
Peterson function fragmentation respectively.  The associated
asymmetry is shown in (c).  The $x_F$ distributions are normalized
to our calculated cross section, Eq.\ (18).  In (a) and (b), the
solid curve is the $\overline \Lambda_c$ distribution (identical to
$\Lambda_c$ for $x_F >0$) while the dashed, dot-dashed, and dotted
curves show $\Lambda_c$ distributions with $r=1$, 10, and 100.  The
predicted model asymmetry for $r=1$ (solid), 10 (dashed), and 100
(dot-dashed) is compared with that from PYTHIA (open circles) in
(c).  At $x_F > 0$, our model predicts no asymmetry.
}
\label{fig5}
\end{figure}
 
The value $r = 1$ is compatible with early low statistics
measurements of charmed baryon production \cite{AgBlam} where equal
numbers of $\Lambda_c$ and $\overline \Lambda_c$ were found in the
range $|x_F|<0.3$. The data is often parameterized as
$(1-|x_F|)^{n_{\Lambda_c}}$, where
\be 
n_{\Lambda_c} = \frac{1-|x_{F,{\rm min}}|}{\langle |x_F| \rangle -
|x_{F,{\rm min}}|} - 2 \, \, . 
\ee
For $x_F < 0$, we predict $n_{\Lambda_c} = 4.6$ for the delta
function and 7.3 for the Peterson function.  The difference is due
to the steeper slope of the fusion model with the Peterson function.
We find rather large values of  $n_{\Lambda_c}$ since the average
$x_F$ is dominated by the leading-twist fusion component at low
$x_F$. If we restrict the integration to $x_F < -0.5$, then
$n_{\Lambda_c}$ decreases to 1.4 independent of the fragmentation
mechanism\footnote{The parameterization $(1-x_F)^n$ is only good if
the distribution is monotonic. However, our two-component
$\Lambda_c$ distribution does not fit this parameterization over all
$x_F$.  At low $x_F$, the leading-twist component dominates.  If
only the high $x_F$ part is included, the value of $\langle |x_F|
\rangle$ is a more accurate reflection of the shape of the intrinsic
charm component.}.  The data on $\Lambda_c$ production measured in
$pp$ collisions at the ISR with $\sqrt{s} = 63$ GeV are consistent
with this prediction.  For $x_F > 0.5$, $n_{\Lambda_c} = 2.1 \pm
0.3$ was found \cite{Chauv} while for  $x_F > 0.35$, $n_{\Lambda_c}
= 2.4 \pm 1.3$ \cite{Zicc}.  Hard charmed baryon distributions have
also been observed at large $x_F$ in $nN$ interactions at the
Serpukhov spectrometer with an average neutron energy of 70 GeV,
$n_{\Lambda_c} = 1.5 \pm 0.5$ for $x_F > 0.5$ \cite{BIS2}. Charmed
hyperons $\Xi_c (usc)$ produced by a 640 GeV neutron beam
\cite{Coteus} do not exhibit a strong leading behavior, $n_{\Xi_c} =
4.7 \pm 2.3$.  This is similar to the delta function prediction for
$n_{\Lambda_c}$ when $x_F < 0$. On the other hand, charmed hyperons
produced with a $\Sigma^-(dds)$ beam \cite{Biagi,WA89} are leading
with $n_{\Xi_c} = 1.7 \pm 0.7$ for $x_F > 0.6$ \cite{Biagi}. Thus in
the proton fragmentation region $r=1$ is compatible with the shape
of the previously measured $\Lambda_c$ $x_F$ distributions. When we
compare the $\Lambda_c$ cross section in the proton fragmentation
region with that of leading-twist fusion, the coalescence mechanism
increases the cross section by a factor of 1.4-1.7 over the fusion
cross section and by 30\%\ over the $\overline \Lambda_c$ cross
section.
 
The extreme value, $r = 100$, was chosen to fit the forward
$\Lambda_c$ production cross section measured at the ISR,
$B\sigma_{pp \rightarrow \Lambda_c X} = 2.84 \pm 0.5$ $\mu$b
\cite{Chauv}, assuming that the charmed quark and $\Lambda_c$ cross
sections are equal, already an obvious overestimate.  This choice
produces a secondary peak in the $\pi^- p$ distributions at $x_F
\sim -0.6$, the average $\Lambda_c$ momentum from coalescence.  Such
a large value of $r$ implies that the intrinsic charm cross section
is considerably larger than the leading-twist cross section.
 
The shape of the distribution with $r=100$ is similar to that due to
diquark coalescence in PYTHIA \cite{PYT}, shown in Fig.\ 1(c), except
that the PYTHIA distribution peaks at $x_F \approx -0.9$ due to the
acceleration induced by the string mechanism.  While a measurement
of the $\Lambda_c$ cross section over the full phase space in the
proton fragmentation region is lacking, especially for $pp$
interactions at $x_F > 0$, no previous measurement shows an increase
in the $\Lambda_c$ $x_F$ distributions as implied by these results. 
However, the reported $\Lambda_c$ production cross sections are
relatively large \cite{Chauv,Zicc,BIS2,Coteus}, between 40 $\mu$b
and 1 mb for $10 \leq \sqrt{s} \leq 63$ GeV.  In particular, the low
energy cross sections are much larger than those reported for the $c
\overline c$ total cross section at the same energy.  This is not
yet understood.
 
A few remarks are in order here.  Some of these experiments
\cite{Zicc,BIS2} extract the total cross section by extrapolating
flat forward $x_F$ distributions back to $x_F = 0$ and also assume
associated production, requiring a model of $\overline D$
production.  On the other hand, the reported $c \overline c$ total
cross sections are usually extracted from $D$ measurements at low to
moderate $x_F$ and would therefore hide any important coalescence
contribution to charmed baryon production at large $x_F$.  High
statistics measurements of charmed mesons and baryons over the full
forward phase space ($x_F > 0$) in $pp$ interactions would help
resolve both the importance of coalescence and the magnitude of the
total $c \overline c$ production cross section.
 
We also chose $r=10$ as an intermediate value.  In this case, a
secondary peak is also predicted but the cross section at $x_F <0$
is only a factor of two to three larger than the fusion cross
section rather than the factor of 21 needed to fit the ISR data at
$x_F > 0.5$.  The magnitude of the second maximum is also less than
the fusion cross section in the central region.
 
An important test of the production mechanism for charm
hadroproduction is the $\Lambda_c$ and $\overline \Lambda_c$
asymmetry, defined as
\be 
{\cal A}_{\Lambda_c/\overline \Lambda_c}  = \frac{d\sigma(\Lambda_c)
- d\sigma(\overline \Lambda_c)}{d\sigma(\Lambda_c) +
d\sigma(\overline \Lambda_c)} \, \, . 
\ee 
If ${\cal A}_{\Lambda_c/\overline \Lambda_c}$ is assumed to arise
only from initial state coalescence, we can estimate the parameter
$r$ from the E791 500 GeV $\pi^- A$ data. Our calculated asymmetries
for the three $r$ values\footnote{We have only shown the delta
function results.  Those with the Peterson function are quite
similar.  The slope increases slightly but the point where ${\cal
A}_{\Lambda_c/\overline \Lambda_c} >0$ does not shift.} are compared
with the results from PYTHIA in Fig.\ 5(c). It is, as expected,
closest to our model with $r=100$ although the asymmetry predicted
by PYTHIA does not increase as abruptly as in our model.
 
Preliminary data from E791 \cite{Kwan} indicate a significant
asymmetry for $x_F$ as small as $-0.1$, albeit with large
uncertainty.  The intrinsic charm model in its simplest form can
only produce such asymmetries if $r \geq 100$, against intuition.
Alternatively, a softer $\Lambda_c$ distribution from coalescence
would make a larger asymmetry at lower $|x_F|$, thus allowing a
smaller $r$. Such a softening could be due to either an important
contribution to $\Lambda_c$ production from a $|n_V g c \overline c
\rangle$ configuration or a different assumption about the $|n_V c
\overline c \rangle$ wavefunction \cite{Navarr}.  In a more
realistic model, both initial and final state coalescence will play
some role in $\Lambda_c$ production.  Final-state coalescence, added
to the leading-twist fusion prediction, would also require a smaller
$r$.
 
At $x_F >0$ there is no asymmetry in $\pi^- p$ interactions since
both the baryon and antibaryon can be produced by fragmentation from
a $|\overline u d c \overline c \rangle$ state and by coalescence
from a $|\overline u d c \overline c q \overline q \rangle$ state
($q = u$,$d$).  Then
\be
\frac{dP^{\Lambda_c}_{\rm ic}}{dx_{\Lambda_c}}  =
\frac{dP^{\overline \Lambda_c}_{\rm ic}}
     {dx_{\overline \Lambda_c}}  =
\frac{dP^F_{\rm ic}}{dx_{\Lambda_c}} \, + \, 
\frac{P_{\rm icq}}{P_{\rm ic}} \,
\frac{dP^C_{\rm ic}}{dx_{\Lambda_c}} \, \, . 
\ee  
The coalescence contribution, obtained from Eq.\ (13), produces a
small shoulder in the distributions at $x_F >0$.  We extract
$n_{\Lambda_c} = 4.1$ for $x_{F, {\rm min}} = 0$, in good agreement
with the NA32 measurement, $n_{\Lambda_c} = 3.5 \pm 0.5$
\cite{Bar2}.  The same mechanism can account for both $\Lambda_c$
and $\overline \Lambda_c$ production in the $\pi^-$ fragmentation
region since no asymmetry is observed \cite{Kwan}, which is also in
accord with the NA32 result, $\sigma(\Lambda_c)/\sigma(\overline
\Lambda_c) \approx 1$ \cite{Bar2}. To look for subtle coalescence
effects as well as to understand the large difference between
$\Lambda_c$ and $\overline \Lambda_c$ production at $x_F \approx
-0.1$ it is important to compare the shape of the momentum
distributions in addition to the asymmetries.
 
Assuming that equal numbers of $D_s$ and $\overline D_s$ are
produced, the $x_F$ distributions are
\be 
\frac{dP^{D_s}_{\rm ic}}{dx_{D_s}}  = \frac{dP^{\overline D_s}_{\rm
ic}}{dx_{\overline D_s}}  =  \frac{dP^F_{\rm ic}}{dx_{D_s}} \, + \,
\frac{P_{\rm ics}}{P_{\rm ic}} \, \frac{dP^C_{\rm ic}}{dx_{D_s}} \,
\, . 
\ee 
The coalescence contribution is given by Eq.\ (14) and the
distributions are shown in Fig.\ 6.  The shoulder at $x_F >0$
predicted in charmed baryon production is absent for $D_s$
production.  Coalescence does not produce a significant enhancement
of $D_s$'s since $P_{\rm ics} < P_{\rm icu}$.  The average momentum
gain over uncorrelated fragmentation of the $|n_V c \overline c
\rangle$ state is small.  The forward distributions in Fig.\ 6 are
only slightly harder than those from leading-twist production. This
is also true for $x_F <0$ where the $D_s$, $\overline D_s$
distributions are not significantly different from the $\overline
\Lambda_c$ distributions even though coalescence is included in the
production of the $D_s$ and not in $\overline \Lambda_c$. When we
compare our distribution with the parameterization
$(1-x_F)^{n_{D_s}}$, we extract $n_{D_s} = 4.7$ at $x_F >0$, in
agreement with the NA32 measurement, $n_{D_s} = 3.9 \pm 0.9$
\cite{Bar2}.
 
\vspace{.5cm}
\begin{figure}[htbp]
\begin{center}
\leavevmode
\epsfbox{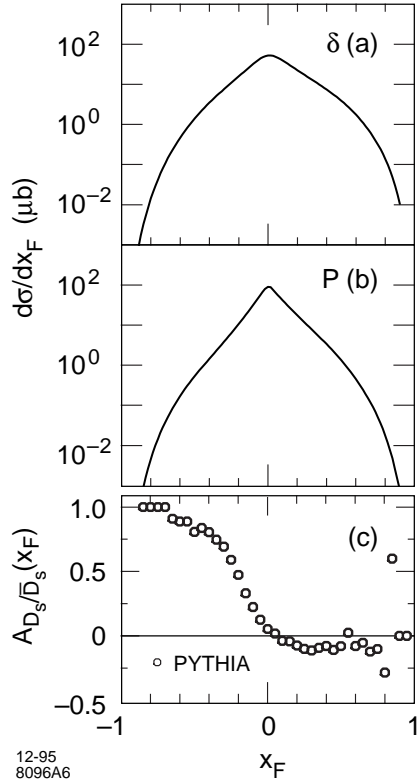}
\end{center}
\caption[*]{The $D_s/\overline D_s$ $x_F$ distributions predicted by
the two-component model in (a) and (b) for delta and Peterson
function fragmentation respectively.  The associated asymmetry from
our model and PYTHIA (open circles) is shown in (c).  Our model
predicts no asymmetry.
}
\label{fig6}
\end{figure}

We find ${\cal A}_{D_s/\overline D_s} (x_F) = 0$ for all $x_F$ in
our model since the production mechanisms are everywhere identical
for the particle and antiparticle. In contrast to the intrinsic
charm model, the $D_s$ excess predicted by PYTHIA in the proton
region, shown in Fig.\ 1(d), leads to a backward asymmetry similar
to ${\cal A}_{\Lambda_c/\overline \Lambda_c}$, shown in Fig.\
6(c).\\[1ex]
 
\noindent
{\bf 4.2 $D \overline D$ production}
\medskip
 
We simplify our discussion of $D \overline D$ pair production by
several respects.  We have assumed that equal numbers of $D^\star$
mesons and, separately, primary $D$ mesons are produced by
fragmentation and that any production excess is the result of
coalescence. Thus the $D \overline D$ and $D^\star
\overline{D^\star}$ $x_F$ distributions are equivalent within the
model and the analysis applies for both types of pairs unless stated
otherwise.  Therefore we shall also implicitly assume that all
secondary $D$'s produced by $D^\star$ decays can be separated from
the primary $D$'s.  We do not consider $D \overline{D^\star}$ or
$D^\star \overline D$ pairs.
 
We use the same pair classification as in our discussion of the pair
distributions from PYTHIA, shown in Fig.\ 2(c) and 2(d).  Then the
probability distributions for intrinsic charm pair production are:
\be  
\frac{dP^{NN}_{\rm ic}}{dx_{D \overline D}} & = &
\frac{dP^{FF}_{\rm ic}}{dx_{D \overline D}} \\
\frac{dP^{NL}_{\rm ic}}{dx_{D \overline D}} & = & 2 \left(
\frac{dP^{FF}_{\rm ic}}{dx_{D \overline D}} + 1.2 \,
\frac{dP^{FC}_{\rm ic}}{dx_{D \overline D}} \right) \\
\frac{dP^{LL}_{\rm ic}}{dx_{D \overline D}} & = &
\frac{dP^{FF}_{\rm ic}}{dx_{D \overline D}} + 1.2 \left(
\frac{dP^{FC}_{\rm ic}}{dx_{D \overline D}} +
P_{\rm ic}^{CC} \, \delta(x_{D \overline D} - 1) +
\frac{P_{\rm icq}}{P_{\rm ic}} \,
\frac{dP^{CC}_{\rm ic}}{dx_{D \overline D}} \right) \, \, . 
\ee
In the above, we have assumed that there is a 20\%\ production
enhancement due to coalescence, {\it e.g.} 
\be 
\frac{dP_{\rm ic}^{D^-}}{dx_{D^-}} = \frac{dP_{\rm ic}^F}{dx_{D^-}}
+ 1.2\frac{dP_{\rm ic}^C}{dx_{D^-}} \, \, . 
\ee 
This assumption is different from our calculation of ${\cal
A}_{D^-/D^+}$ in \cite{VB} where we assumed that the $D^-$ and $D^+$
production probabilities were equal even though the $D^-$ is
produced by coalescence and the $D^+$ is not. This resulted in a
small negative asymmetry at low $x_F$.  When ${\cal A}_{D^-/D^+}$ is
recalculated with the $D^-$ distribution in Eq.\ (28), the model
asymmetry is never negative and better agreement with the data
\cite{Carter,WA82,E7692} is found.
 
A factor of two has been included in the $NL$ distribution because
there are two sources of the $NL$ pairs ($D^-D^+$ and
$D^0\overline{D^0}$) relative to the $LL$ and $NN$ pairs.  This
factor is also included in the leading-twist production cross
section.  The last term in the $LL$ distribution, from the
six-particle Fock state, is the only significant source of pairs due
to double coalescence, causing the $LL$ distributions to be somewhat
harder than the $NL$ distributions. The pair distributions are given
in Fig.\ 7(a) and 7(b).  Note that even the $NN$ distribution with
the Peterson function in Fig.\ 7(b) is broadened considerably over
the leading twist distribution shown in Fig.\ 2(b). The predicted
slopes of the pair distributions from the various sources begin to
differ for $x_F > 0.25$.
 
\vspace{.5cm}
\begin{figure}[htbp]
\begin{center}
\leavevmode
\epsfbox{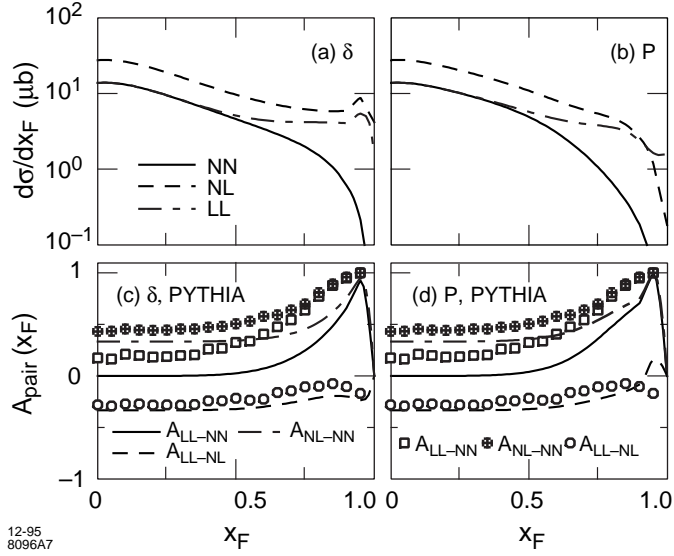}
\end{center}
\caption[*]{The $D \overline D$ pair distributions 
from $\pi^-p$ interactions at 500 GeV for $NN$ (solid),
$NL$ (dashed), and $LL$ (dot-dashed) pairs are shown in (a) and (b)
for delta function and Peterson function fragmentation.  The
intrinsic charm model distributions are given in Eqs.\ (25)-(27). 
The asymmetries for delta function and Peterson function
fragmentation from the two-component model, calculated using Eqs.\
(29)-(31), are shown in (c) and (d).  They are compared with the
$D^\star \overline D^\star$ asymmetries from PYTHIA for ${\cal
A}_{LL-NN}$ (solid curve and squares), ${\cal A}_{LL-NL}$ (dashed
curve and circles), and ${\cal A}_{NL-NN}$ (dot-dashed curve and
crosses).
}
\label{fig7}
\end{figure}

We define the following three asymmetries:
\be 
{\cal A}_{LL-NN} & = &
\frac{d\sigma(LL) - d\sigma(NN)}{d\sigma(LL) + d\sigma(NN)} \\
{\cal A}_{LL-NL} & = &
\frac{d\sigma(LL) - d\sigma(NL)}{d\sigma(LL) + d\sigma(NL)} \\
{\cal A}_{NL-NN} & = &
\frac{d\sigma(NL) - d\sigma(NN)}{d\sigma(NL) + d\sigma(NN)}\,\, , 
\ee
shown in Fig.\ 7(c) and 7(d) for delta and Peterson function
fragmentation. The corresponding $D^\star \overline{D^\star}$
asymmetries from PYTHIA are also shown since the leading assignments
are unaffected by decays. It is interesting to study all three
asymmetries.  Since twice as many $NL$ pairs are produced by
definition, ${\cal A}_{LL-NL}$ is negative for all $x_F$.  The
change in ${\cal A}_{LL-NL}$ at $x_F \sim 1$ in Fig.\ 7(d) is due to
double coalescence from the $|\overline u d c \overline c \rangle$
state.  Larger values of ${\cal A}_{NL-NN}$ and ${\cal A}_{LL-NN}$
are predicted by PYTHIA than by our model due to our different
assumptions about particle production.  However, the general trends
are quite similar for the final-state coalescence mechanism of
PYTHIA and the initial-state coalescence of our model. This is
perhaps not surprising since the single $D$ asymmetries from the two
models would be quite similar if the same assumptions were made
about the initial production ratios at $x_F = 0$.\\[3ex]
 
\begin{center}
{\bf 5. Conclusions}
\end{center}
 
We have studied single charmed hadron and charmed meson pair
longitudinal momentum distributions and the related asymmetries
within the intrinsic charm model and PYTHIA.  In conventional
leading-twist perturbative QCD, there is no asymmetry between
charmed and anticharmed hadrons.  In the intrinsic charm model as
well as in PYTHIA, the asymmetry is clearly due to coalescence.
However, we find that the asymmetries alone cannot tell the full
story, especially at $x_F >0$.  The individual $x_F$ distributions
are needed over all $x_F$ to unravel the production properties of
the charmed hadrons since only the shapes of these distributions can
reveal deviations from the fusion predictions.
 
The $D \overline D$ pair asymmetries are also quite interesting,
particularly at high pair momentum.  We have not considered their
production at negative $x_F$ in this paper due to the ambiguity in
leading particle assignments.  However, since only $D^-$ or
$\overline{D^0}$ can be produced by coalescence from the
five-particle Fock state, a study of $\Lambda_c \overline{D^0}$ or
$\Sigma_c D^-$ pairs could prove more enlightening.
 
Given our $\Lambda_c$ predictions at $x_F <0$ in $\pi^- p$
interactions, it would be quite interesting if high statistics
measurements of $\Lambda_c$ production can be made in $pp$
collisions at $|x_F| >0$ to clarify charmed baryon production. 
Since we would predict similar behavior for bottom hadron
production, such studies would also be of interest.  Although a rise
in $d\sigma/dx_F$ with $x_F$ seems counterintuitive, such
distributions have been observed in diffractive $\Xi^-$ and
$\Omega^-$ production in $\Xi^-$Be interactions \cite{Sigma}. 
Studies of charmed baryon production by a hyperon beam within the
context of this model are underway.
 
\begin{center}
{\bf Acknowledgments}
\end{center}

\medskip 
We would like to thank J. A. Appel, T. Carter, W. Geist, P. Hoyer,
J. Leslie, S. Kwan, M. Moinster,  A. Mueller, E. Quack, T. 
Sj\"{o}strand, and W.- K. Tang for discussions. This work was
supported in part by the Director, Office of Energy Research,
Division of Nuclear Physics of the Office of High Energy and Nuclear
Physics of the U. S. Department of Energy under Contract No.
DE-AC03-76SF0098 and DE-AC03-76SF00515.

\end{document}